\documentclass{revtex4}

\newcommand{\be}{\begin{equation}}
\newcommand{\ee}{\end{equation}}
\newcommand{\<}{\langle}
\renewcommand{\>}{\rangle}

\newcommand{ \K }{ \raisebox{-0.8ex} {\scriptsize \it K} }

\begin{document}

\title{Bosonization and IBM \footnote{\uppercase{T}his work is supported 
\uppercase{EEC} under the contract \uppercase{HPRN-CT-2000-00131}.}}

\author{Fabrizio Palumbo}

\address{INFN -- Laboratori Nazionali di Frascati, \\ 
P.~O.~Box 13, I-00044 Frascati, ITALIA\\
E-mail: fabrizio.palumbo@lnf.infn.it}

\begin{abstract}
We derive a boson Hamiltonian  from 
a Nuclear Hamiltonian whose potential is expanded in pairing multipoles 
and determine the  fermion-boson mapping of operators. We use a  new method of 
bosonization based on the  evaluation of the partition  function restricted to 
the bosonic composites of interest. By rewriting  the partition function so obtained 
in functional form we get the euclidean action of the composite  bosons from which 
we can derive the Hamiltonian. Such a procedure respects all the fermion symmetries.
\end{abstract}

\maketitle

\section{Introduction}

The IBM of Arima and Iachello~\cite{Iach}, is most successful in describing the low energy nuclear excitations. The bosons of this model are understood as virtual pairs of nucleons, analogous to the Cooper pairs of superconductivity~\cite{Namb}. But no general procedure to reformulate the nuclear theory in
terms of the effective bosonic degrees  of freedom has been found.

 The first attempt in this direction has been performed, as far as we know, by
Beliaev and Zelevinsky~\cite{Beli}. But this work  makes use of the Bogoliubov transformation which violates nucleon
number conservation. Moreover the bosonization is achieved only within a perturbation scheme. 

 The first work which relates the IBM to a nucleon Hamiltonian is due
to Otsuka, Arima and Iachello~\cite{Otsu}. These authors got exact results for the pairing interaction
in a single j-shell. Their result was somewhat generalized~\cite{Otsu1}, but a full solution of the
problem has not yet been achieved. 

There are several recipes for bosonization~\cite{Klei}, mostly based on the idea of mapping a fermion model
space into a boson space. This requires a truncation of the nucleon  space whose effect is in general not easy to
control. 

In order to avoid the limitations of previous works we try a different approach
where we do not  assume  any property of the composites, other than their dominance at low energy. In particular
their structure will be determined only at the end of the calculation. The problem of 
truncation of the nucleon space will then be traded by the problem
of decoupling some bosons from the others, but in a setting where one can hopefully have a better control.

To implement Boson Dominance we perform a functional evaluation of the partition function restricted to
boson composites. In this way we  get the euclidean action of these composites  and their coupling
to external fields in closed form. All the fermion symmetries, in particular fermion number conservation, are respected.
 The  bosonization is therefore achieved in the path integral formalism, and all
physical quantities can be evaluated by standard methods. The first step, necessary also in the derivation of the
Hamiltonian, is to find the minimum of the action at constant fields. Depending on the solution, one has spherical or
deformed nuclei. In the latter case rotational excitations appear as Goldstone modes associated to the spontaneous
breaking of rotational symmetry. The notion of spontaneous symmetry breaking survives in fact with a precise definition
also in finite systems~\cite{Barb}. We want to emphasize that the  closed form of the action opens the way to numerical
simulations of fermionic systems in terms of bosonic variables, avoiding the "sign problem".  

To compare with the IBM we can either write the path integral of the latter, or derive the Hamiltonian corresponding
to our action. We will make here the second choice. But to derive the Hamiltonian we must perform an
expansion in the inverse of the shell degeneracy.

Bosonization appears in several many-fermion systems and relativistic field theories. The effective bosons fall into
two categories, depending on their fermion number. The Cooper pairs of the BCS model of superconductivity, of the
IBM of Nuclear Physics, of the  Hubbard model of high $T_c$ superconductivity~\cite{Cini} and of color superconductivity
in QCD have fermion number 2. Similar  composite bosons with fermion number zero appear as phonons, spin waves and
chiral mesons in QCD. They can be included in the present formalism by replacing in the composites one fermion operator
by an antifermion (hole) one. Indeed the approach we are going to present can be applied, as far as we can see without
any conceptual difficulty, in all the above cases, as it has been argued in a brief report of the method~\cite{Palu}.

A different approach to bosonization which also avoids any mapping is based on the Hubbard-Stratonovich
transformation. The latter renders quadratic the fermionic interaction by introducing bosonic auxiliary fields which
in the end become the physical fields. The typical resulting structure is that of chiral theories~\cite{Mira}. In such
an approach an energy scale emerges naturally, and only excitations  of lower energy can be described by the auxiliary
fields~\cite{Barb}. At present the relation with the present approach has not been fully clarified.

The paper is organized in the following way. In Sec. 2 we define coherent states of composites and their
properties. In Sec. 3 we derive the path integral for the composites and find the effective bosonic action.
We restrict ourselves for simplicity to a  nuclear interaction given as a sum of pairing multipoles, but
more general forces can easily be included and will be discussed in future works. The effective action we derive 
is, apart from the above limitation, general. In Sec. 4 we restrict ourselves to a single j-shell  with pairing multipoles and in Sec. 5 we determine the
corresponding Hamiltonian. In  Sec. 6 we report our conclusions.

\section{Coherent states of composites}

Composites of fermion number 2 are defined in terms of the fermion creation operators 
${\hat c}^{\dagger}$
\be
{\hat b}^{\dagger}_J= { 1\over 2 } {\hat c}^{\dagger} B^{\dagger}_J {\hat c}^{\dagger}
= { 1\over 2 } \sum_{m_1,m_2}{\hat c}^{\dagger}_{m_1}\left( B^{\dagger}_J\right)_{m_1,m_2}
 {\hat c}^{\dagger}_{m_2}.
\ee
 In the above equation the $m$'s represent all the fermion intrinsic quantum numbers and position 
coordinates and $J$ the corresponding labels of the composites. Composites of fermion number zero can be
obtained by replacing one of the fermion operators by an antifermion one.
The structure matrices $B_J$ have dimension $ 2 \Omega$ independent of $J$. Their form is determined
by the fermion interaction as explained in the sequel, but we assume that they will satisfy the
relations 
\be
\mbox{tr} (B^{\dagger}_J \,  B_K ) = 2 \,  \delta_{J,K}.  \label{norm}
\ee
 We also assume them to be
nonsingular. Then their dimension is twice the index of nilpotency of the composites, which is the largest
integer $\nu$ such that $ \left( \hat{b}_J \right)^{\nu} \neq 0 $. It is
obvious that a necessary condition for a composite to resemble a boson, is that its index of nilpotency be
large. But this condition in general is not sufficient, and we must require also that
\be
\det ( \Omega B^{\dagger}B)^n \sim 1.  \label{suff}
\ee
 A convenient way to get the euclidean path integral from the trace of the transfer
matrix is to use coherent states\cite{Nege}. If we are interested in states with $n= \overline{n} + \nu$
bosons for an arbitrary reference number $ \overline{n}$  we introduce the  operator
\be 
{\mathcal P}_{\overline{n}} = { ( \Omega - {\overline n})^2 \over \Omega^2}
\int db^* db \< b| b \>^{-1}|b \> \< b|
\ee
constructed in terms of coherent states of composites
\be
|b \> = |\exp \left( \sum_J b_J {\hat b}^{\dagger}_J \right) \>.
\ee
  We would like it to be the identity in the fermion subspace of the composites. Let us see its action on 
composite operators. Let us first consider the case where there is only one composite with structure
function satisfying the equation
\be
B^{\dagger} B= { 1\over \Omega} 1\!\!1.  \label{Bzero}
\ee
In order to evaluate the matrix element $\<b_t|b_{t-1}\>$ we introduce  between the bra and the ket the
identity in the fermion Fock space
\be
{\mathcal I} = \int dc^* dc \<c | c\>^{-1} | \exp (- c^* {\hat c} ) \> \< \exp (- c \, {\hat c}^{\dagger})|
\ee
where the $c^*,c$ are Grassmann variables. We thus find
\begin{eqnarray}
\<b_t| b_{t-1}\> &=& \int dc^* dc \, E(c^*,c,b_t^*,b_{t-1}) = \left(1+ {1\over \Omega} b_t^* b_{t-1}
\right) ^{\Omega}
\end{eqnarray}
where
\begin{eqnarray}
E(c^*,c,b^*,b)  & = & \exp \left( -c^*c + {1\over 2 } b^* \, c \, B \, c + 
{ 1\over 2} b  \, c^* B^{\dagger} c^* \right).
\label{E}
\end{eqnarray}
Therefore the action of ${\mathcal P}_{\overline{n}}$ on the composites 
\be
|{\mathcal P}_{{\overline n}} \, ({\hat b}^{\dagger})^n \>= 
\left( 1 - { \nu \over \Omega - {\overline n}} \right)^{-1}
\left( 1 - { \nu +1 \over \Omega - {\overline n}} \right)^{-1}  |({\hat b}^{\dagger})^n \>
\ee
shows that it behaves like the identity in the neighborhood of the reference state up to an error of order $
\nu /(\Omega - {\overline n})$, namely the measure $\<b|b\>^{-1}$ is essentially uniform with respect to any
reference state.

 It is  worth while noticing that in the limit of infinite $\Omega$ we recover exactly the expressions valid
for elementary bosons, in particular
\be
\<b_t|b_{t-1}\>= \left(1+ {1\over \Omega} b_t^* b_{t-1} \right) ^{\Omega} \rightarrow \exp (b_t^*
b_{t-1}),\,\,\,
\Omega \rightarrow \infty.
\ee

In the general case of many composites the above equations become
\be
\<b_t|b_{t-1}\> = \left[ \det \left( 1\!\!1 + \beta_t^* \beta_{t-1} \right) \right]^{1 \over 2},
\ee
where $\beta_t^*  = \sum_J (b_J)_t^*  B_J $.
Then using the condition~\ref{suff} we find again that ${\mathcal P}$ approximates the identity with an error  of
order
$ 1 /
\Omega$
\be
{\mathcal P}| ({\hat b_{I_0}}^{\dagger})^{n_0}...{\hat b_{I_i}}^{\dagger})^{n_i}\> =
| \left( ({\hat b_{I_0}}^{\dagger})^{n_0}...{\hat b_{I_i}}^{\dagger})^{n_i} \, + O(1 / \Omega) \right)\>.
\ee
Identifying the operator $ {\mathcal P}$ with the identity in the subspace of the composites is the only
approximation we will make in the derivation of the effective boson action.

\section{Composites path integral}

Now we are equipped to realize our program. The 
first step is the evaluation of the partition function $Z_c$ restricted to fermionic composites. To
this end we divide the inverse temperature in $N_0$ intervals of spacing $\tau$
\be
\tau = { 1 \over N_0 T}  
\ee
and write
\be
Z_c = \mbox{tr} \left( {\mathcal P} \exp \left( - H \tau \right) \right)^{N_0}.
\ee
 We will restrict ourselves to a
Hamiltonian with interactions which can be written as a sum of pairing multipoles
\begin{eqnarray}
{\hat H} &=& {\hat c}^{\dagger} h_0 \, {\hat c} -
\sum_K g\K  \, { 1\over 2} \, {\hat c}^{\dagger} F_K^{\dagger} {\hat c}^{\dagger} \, 
{ 1\over 2}\, {\hat c}\, F_K \, {\hat c}. 
\end{eqnarray}
 The single particle term includes the single
particle energy with matrix $e$, any single particle interaction with external fields described by the matrix
${\mathcal M}$ and the chemical potential $\mu$
\be
h_0= e + {\mathcal M} - \mu.
\ee
Therefore we will be able to solve the  problem of fermion-boson mapping by determining the
interaction of the composite bosons with external fields. We assume for the potential form factors the
normalization
\be
 \mbox{tr} ( F_K^{\dagger} F_K) = 2 \, \Omega .
\ee
 For the following manipulations we need the Hamiltonian in antinormal form
\be
{\hat H} = H_0 - {\hat c} \,  h^T {\hat c}^{\dagger} -
\sum_K g\K \,  {1\over 2} \, {\hat c} F_K {\hat c}
\, { 1\over 2} \, {\hat c}^{\dagger} F^{\dagger}_K {\hat c}^{\dagger}
\ee
where the upper script $T$  means "transposed" and
\be
h = h_0 -  \sum_K g\K   F_K^{\dagger} F_K,\,\,\,
H_0 = { 1\over 2} \mbox{tr} ( h + h_0 ).
\ee
Now we must evaluate the matrix element $\<b_t| \exp(- \tau {\hat H}) |b_{t-1} \>$. To this end we expand to 
first order in
$\tau$ (which does not give any error in the final $\tau \rightarrow 0$ limit) and
insert the operator ${\mathcal P}$ between annihilation and creation operators
\begin{eqnarray}
& &\< b_t| \exp ( - \tau \hat{H} ) |b_{t-1}\> = \exp( - H_0 \tau) \<b_1| {\mathcal{P}} -  \hat{c} \,  h^T \tau \, {\mathcal{P}}
\hat{c}^{\dagger}
\nonumber\\
& & \,\,\,\,\,  \times \sum_k g_k \tau  \, {1\over 2} \, {\hat c} F_K {\hat c} \, {\mathcal{P}}
\, { 1\over 2} \, {\hat c}^{\dagger} F^{\dagger}_K {\hat c}^{\dagger} |b\>.
\end{eqnarray} 
Using the identity in the fermion Fock space we find
\begin{eqnarray}
& &\<b_t|  \exp (- \tau \hat{H} ) |b_{t-1}\>^{-1} =  \int dc^* dc  \, E(c^*,c,b_t^*,b_{t-1})
\nonumber\\
& & \,\,\,\,\,\,\,\,\,\, \times   \exp(- H_0 \tau   - c^* h \, \tau c )
\exp \left(   \sum_K g_K \tau \, {1\over 2} c \,F_K \,c  
 \, { 1\over 2}  c^* F_K^{\dagger} c^* \right)
\end{eqnarray}
where the function $E(c^*,c,b^*,b)$ is defined in~(\ref{E}).
By means of the Hubbard-Stratonovich transformation we can make the exponents quadratic in the Grassmann
variables and evaluate the Berezin integral
\begin{eqnarray}
& & \<b_t| \exp(- \tau {\hat H}) |b_{t-1} \> = \det R \, \exp (- H_0 \tau ) \int \prod_K da_K^* da_K
\nonumber\\
& & \,\,\, \times  \exp (   - a^* \cdot a  ) \exp \left\{ 
 { 1\over 2} \mbox{tr} \ln \left[ 1\!\!1 + \left( \beta_t^* + \sum_{K_1} \sqrt{ g_{K_1}
\tau} 
\,
 a^*_{K_1} F_{K_1}\right)
\right. \right.
\nonumber\\
& & \,\,\, \left. \left.  \times R^{-1} \left( \beta_{t-1} + \sum_{K_2} \sqrt{ g_{K_2} \tau} 
\, a_{K_2} (F_{K_2})^{\dagger}\right) (R^T)^{-1} \right] \right\}
,
\end{eqnarray}
where $
R = 1\!\!1 + h \, \tau.
$
Setting
$
\Gamma_t= \left( 1\!\!1 + \beta_t^* \beta_{t-1}\right)^{-1}
$
and performing the integral over the auxiliary fields $a\K^*,a\K$ we get the final expression of the euclidean
action 
\begin{eqnarray}
& & S(b^*,b) = \tau \sum_t \left\{ H_0  -\mbox{tr} \, h + { 1\over 2 \tau} \mbox{tr} [ \ln (1\!\!1 + \beta_t^* \beta_t) 
+ \ln \Gamma_t]  \right.
\nonumber\\
& & \left. - { 1\over 4}  \sum_K g_K  \left[   \mbox{tr} (\Gamma_t \beta_t^*  F_K^{\dagger}) \,
\mbox{tr}(\Gamma_t F_K \beta_{t-1}) +  2 \, \mbox{tr} \left( \Gamma_t  F^{\dagger}_K F_K \right) \right. \right.
\nonumber\\
& & \left. \left. 
- \mbox{tr} [ \Gamma_t \beta_t^* F_K^{\dagger}, \Gamma_t  F_K \beta_{t-1}]_+   
  \right] \right.
 \left.  + { 1\over 2}  \mbox{tr} \left[ \Gamma_t  \beta_t^*  
\, ( \beta_{t-1} \, h^T + h \,\, \beta_{t-1} ) \right] \right\}  \label{bosaction}  
\end{eqnarray}
where $[..,..]_+$ is an anticommutator.
This action differs from that of elementary bosons because

i) the time derivative terms are non canonical. Indeed expanding the logarithms we get
\begin{eqnarray}
& &\mbox{tr} \, [ \, \ln (1\!\!1 + \beta_t^* \beta_t) + \ln \Gamma_t  \, ]  = { 1\over 2 }
\left\{ \beta_t^* \nabla_t \beta_t - { 1\over 2}
[(\beta_t^* )^2 \nabla_t (\beta_t^* )^2 ] +...\right\},
\end{eqnarray}
where
$
\nabla_t \, f = { 1 \over \tau} \left( f_{t+1} - f_t \right).
$
The first term is the canonical one, while the others contain the derivative of powers of the boson variables.
The canonical form of the first term is due to the normalization of Eq.(\ref{norm}) of the structure functions,
otherwise 
$\beta_t$ and $\beta_{t-1}$ would not have the same coefficient. Note the difference of the noncanonical terms
with respect to the chiral expansions, where there are powers of derivatives, rather than derivatives of powers.

ii) the coupling of the chemical potential (which appears in $h$) is also noncanonical. Indeed expanding $\Gamma_t$
we get $
\mu  \, \mbox{tr} \left( \beta_t^* \beta_{t-1} - \beta_t^* \beta_t \beta_t^* \beta_{t-1}+...\right),
$
and only the first term is canonical

iii) the function $\Gamma$ becomes singular when the number of bosons is of order $\Omega$, as it will become clear in
the sequel. This reflects the Pauli principle.

We remind the reader that the only approximation done concerns the operator ${\mathcal P}$. Therefore
these are to be regarded as true features of compositeness.

The bosonization of the system we considered has thus been accomplished. In particular the fermionic
interactions with external fields can  be expressed in terms of the bosonic terms which involve the matrix
${\mathcal M}$ (appearing in $h$). The dynamical problem of the interacting (composite) bosons can be solved
within the path integral formalism. This includes the new interesting  possibility of a
numerical simulation of the partition function which could now be  performed with bosonic variables 
avoiding the sign problem.

 Part of the dynamical problem is the 
determination of the structure matrices $B_J$. This can be done by expressing the energies in terms of
them and applying a variational principle which gives rise to an eigenvalue equation.

\section{The action in a single $j$-shell}  

In this paper we restrict ourselves to a system of nucleons of in a single $j$-shell. Then we identify the quantum number $K$ with the
boson angular momentum, $ K = (I_K,M_K)$, so that the form  factors of the potential are proportional to Clebsh-Gordan
coefficients
\be
(F_{IM})_{m_1,m_2}= \sqrt{2 \Omega} \, C_{j m_1 j m_2}^{ IM}, \,\,\, \Omega = j + { 1\over 2}. 
\ee
In such a case the structure matrices are completely determined 
by the angular momentum of the composites and the normalization conditions (\ref{norm})
$
B_J= \Omega^{- { 1 \over 2}} F_J.
$
  The
points i) and ii) following Eq.~(\ref{bosaction}) are the only difficulties in the derivation of the Hamiltonian which
could be otherwise read from the action. We
can overcome them by performing an expansion in inverse powers of $\Omega$. We will retain only the first order
corrections, which are of order $\Omega^0$,  with the exception of the coupling with external fields where they are of
order
$\Omega^{-1}$. In this approximation the first difficulty is overcome because
noncanonical time derivatives are of order $1 / \Omega$ and the second one because the only noncanonical
coupling of the chemical potential of order $ \Omega^0$ comes from the only term of the chemical potential of
order $\Omega$, which can be shown to be $ \mu \sim - { 1 \over 2} \,  g_0 \, \Omega$, independent of the number of bosons.

The resulting action is
\begin{eqnarray}
& & S(b^*,b)  = \sum_t \left\{ \sum_{K_1 K_2} b^*_{K_1} \left[ ( \nabla_t - 2 \mu )  +
\omega \right]_{K_1 K_2}  b_{K_2} 
\right.
\nonumber\\ 
 & &  \left.  +  \sum_{I_1 I_2 I_3 I_4}  \sum_{IM}  W^I_{I_1 I_2 I_3 I_4} \left( b^*_{I_1} \,
b^*_{I_2}\right)_{IM}\left( b_{I_3} \, b_{I_4}\right)_{IM} \right\}
\nonumber\\
\end{eqnarray}
where all the $b^*$'s and all the $b$'s are
at time $t,t-1$ respectively, and
\begin{eqnarray}
\omega_{K_1 K_2} &=& {1 \over \Omega} \mbox{tr} \left(\, F_{K_1} F^{\dagger}_{K_2} \, e \right)
 -g_{I_1} \Omega \, \delta_{K_1 K_2}
\nonumber\\
 W^I_{I_1I_2I_3I_4} &=&  \left( - 2 g_0  + \sum_{i=1}^4 g_{I_i}  \right)  \Pi_{i=1}^4 [(2I_i+1)]^{1/2}  
\Omega 
  \left\{
    \begin{array}{ccc}
       j & j & I_1   \\
       j & j & I_2   \\
       I_3 & I_4 & I \\
    \end{array}
     \right\} 
\nonumber\\
\left( b_{I_3} \,  b_{I_4}\right)_{IM} & = &\sum_{M_3 ,M_4} C_{I_3,M_3,I_4,M_4}^{I,M} \, 
b _{I_3 M_3} b_{I_4 M_4}.
\end{eqnarray}
Notice the factor 2 in front of the chemical potential due to the fact that the composites have fermion number
2.

\section{The Hamiltonian}

The Hamiltonian is obtained\cite{Nege} by omitting the time derivative  and  chemical potential terms,
and replacing the variables $b^*,b $ by corresponding creation-annihilation operators
$\hat{a}^{\dagger},
\hat{a}$, satisfying canonical commutation relations
\begin{eqnarray}
 \hat{H} &=& \sum_{I_1M_1I_2M_2} \omega_{I_1M_1I_2M_2} \hat{a}^{\dagger}_{I_1M_1}  \hat{a}_{I_2M_2}  + 
\sum_{I_1I_2I_3I_4} 
\sum_{IM} 
\nonumber\\ 
 & &  \left\{ W^I_{I_1I_2I_3I_4} \left( \hat{a}^{\dagger}_{I_1} \,
\hat{a}^{\dagger}_{I_2}\right)_{IM}\left( \hat{a}_{I_3} \, \hat{a}_{I_4}\right)_{IM} \right\}.
\end{eqnarray}
It is easy to check that, due to the symmetries of the 9j symbols, it is hermitian.

From the interaction with external fields we get the fermion-boson mapping of other operators
\begin{eqnarray}
 \hat{c}^{\dagger} {\mathcal M} \hat{c} & & \rightarrow  \sum_{I_1M_1I_2M_2}  { 2 \over  \Omega}
\mbox{tr} \left( F_{I_1M_1} {\mathcal M} \, F_{I_2M_2}^{\dagger} \right)  \hat{a}^{\dagger}_{I_1M_1} \hat{a}_{I_2M_2} 
\nonumber\\
& &
 + \sum_{\mbox{all} \, I,M} \left({ 2 \over  \Omega}\right)^2 
\mbox{tr} \left( F_{I_1M_1} {\mathcal M} \, F_{I_4M_4}^{\dagger} F_{I_2M_2} F_{I_3M_3}^{\dagger}\right) 
\nonumber\\
& & 
 \times \hat{a}^{\dagger}_{I_1M_1} \hat{a}^{\dagger}_{I_2M_2} \hat{a}_{I_3M_3} \hat{a}_{I_4M_4}.
\end{eqnarray}
 We remind the reader that the above Hamiltonian has been derived under the condition $n<< \Omega$ in a single
subshell. 
Therefore if we further  assume 
$
e_{m_1 m_2} = \overline{e} \, \delta_{m_1 m_2},
$
the single boson energy matrix is diagonal
$
\omega_{I_1 I_2} = ( 2 \, \overline{e} - g_{I_1} \Omega) \delta_{I_1 I_2}.
$
 But the bosonic interactions couple all the bosons with angular momenta for which the $9j$ symbols do not vanish,
even if the corresponding potentials do vanish.

\section{Summary}

 We have developed a general approach to the problem of bosonization where we introduce fermionic composites
without any preliminary mapping of the fermion model space into a bosonic one. Restricting the trace
in the partition function of the system to the composites we get the euclidean action of the effective
bosons in closed form. The only approximation made concerns the identity operator in the space of the composites.

It is perhaps worth while to spend a few words about the nature of this approximation. Indeed it might appear that
two are the approximations involved. The first one is the restriction of the partition function to composites.
This is the fundamental physical assumption of Boson Dominance. Then we replace the identity in the composite subspace by the operator
${\mathcal P}$, which seems a further approximation. But ${\mathcal P}$ differs appreciably from the identity only for states
with many bosons, states which cannot resemble elementary bosons because of the Pauli principle. We therefore deem
that the two approximations are essentially one and the same.
 
  The nuclear dynamics  can  be studied by the methods of path integrals, including numerical simulations which now are
not affected by the sign problem.

  To derive the Hamiltonian of the IBM we must make recourse to an expansion in the inverse of the index of nilpotency
of the composites. In the present work we restricted ourselves to a nucleon model space of a single j-shell and to
a number of bosons much smaller than the index of nilpotency. Both limitations can easily be removed. Concerning the
second one, some care must be exercized to respect particle number conservation, as done for instance in~\cite{Barb}.
The first one requires a parametrization of the structure functions according to
\be
(B_{J,M})_{m_1,m_2}= \sum_{j_1 j_2} p_{J j_1 j_2} C^{JM}_{j_1 m_1 j_2 m_2}.
\ee
Now the energies of the bosons are functions of the parameters $p$. A variational principle applied to these
energies generates an equation for these parameters. The solution to this equation can in general be found only 
numerically, but the Hamiltonian and the other operators retain their analytic expressions.

\end{document}